\documentclass[fleqn,usenatbib]{mnras}

\usepackage{newtxtext,newtxmath}
\usepackage[T1]{fontenc}

\DeclareRobustCommand{\VAN}[3]{#2}
\let\VANthebibliography\thebibliography
\def\thebibliography{\DeclareRobustCommand{\VAN}[3]{##3}\VANthebibliography}

\usepackage{graphicx}	% Including figure files
\usepackage{amsmath}	% Advanced maths commands
\usepackage{xcolor} %color text

\newcommand{\kel}{\,K} % Kelvin
\newcommand{\per}{\,\%}%percentage
\newcommand{\mbar}{\,mbar} %mbar
\newcommand{\um}{\,µm} %microns

\title[The fate of icy pebbles undergoing sublimation]{The fate of icy pebbles undergoing sublimation in \\protoplanetary discs}

\author[S.Spadaccia et al.]{
Stefano Spadaccia,$^{1}$\thanks{E-mail: stefano.spadaccia@space.unibe.ch}
Holly L. Capelo,$^{1}$
Antoine Pommerol,$^{1}$
Philipp Schuetz,$^{2}$
Yann Alibert,$^{1}$
\newauthor{
Katrin Ros$^{3}$
and Nicolas Thomas$^{1}$
}
\\
$^{1}$ Space Research and Planetary Sciences Division, Physikalisches Institut, University of Bern, Sidlerstrasse 5, 3012 Bern, Switzerland \\
$^{2}$Lucerne University of Applied Sciences and Arts, Technikumstrasse 21, 6048 Horw, Switzerland\\
$^{3}$Lund Observatory, Department of Astronomy and Theoretical Physics, Lund University, Box 43, 221 00 Lund, Sweden
}

\date{Accepted XXX. Received YYY; in original form ZZZ}

\pubyear{2021}

\begin{document}
\label{firstpage}
\pagerange{\pageref{firstpage}--\pageref{lastpage}}
\maketitle

\begin{abstract}
Icy pebbles may play an important role in planet formation close to the water ice line of protoplanetary discs. There, dust coagulation is more efficient and re-condensation of vapor on pebbles may enhance their growth outside the ice line. Previous theoretical studies showed that disruption of icy pebbles due to sublimation increases the growth rate of pebbles inside and outside the ice line, by freeing small silicate particles back in the dust reservoir of the disc. However, since planet accretion is dependent on the Stokes number of the accreting pebbles, the growth of planetesimals could be enhanced downstream of the ice line if pebbles are not disrupting upon sublimation. We developed two experimental models of icy pebbles using different silicate dusts, and we exposed them to low-temperature and low-pressure conditions in a vacuum chamber. Increasing the temperature inside the chamber, we studied the conditions for which pebbles are preserved through sublimation without disrupting. We find that small silicate particles ($<50$\um) and a small quantity of ice (around $15$\per\, pebble mass) are optimal conditions for preserving pebbles through sublimation. Furthermore, pebbles with coarse dust distribution ($100-300$\um) do not disrupt if a small percentage ($10-20$\per\, mass) of dust grains are smaller than $50$\um. Our findings highlight how sublimation is not necessarily causing disruption, and that pebbles seem to survive fast sublimation processes effectively. 
\end{abstract}


%%%%%%%%%%%%%%%%%%%%%%%%%%%%%%%%%%%%%%%%%%%%%%%%%%

%%%%%%%%%%%%%%%%% BODY OF PAPER %%%%%%%%%%%%%%%%%%

\section{Introduction} \label{Introduction}

Protoplanetary discs are collapsed clouds of gas and dust surrounding young stars. From micrometer-sized dust particles, planetary embryos form and a completely new planetary system appears eventually. The direct growth of dust up to the size of planetesimals ($1-100$\,km) is inhibited. Particles bigger than mm-size tend to disrupt in collisions, and boulder-size objects interact heavily with the gas in the disc, losing angular momentum and drifting toward the central star \citep{weidenschilling_origin_1997, blum_growth_2008, birnstiel_simple_2012}. Streaming instability \citep{youdin_streaming_2005,johansen_protoplanetary_2007} is a promising mechanism to overcome this inhibition and actually form planetesimals: mm-cm size aggregates (referred hereafter as `pebbles'), when numerous enough, concentrate in filaments in the disc, these latter eventually collapsing to form planetesimals.

In this scenario, ice lines in the disc can play a key role in planet formation. The ice line of a certain volatile species in the disc is the location where the temperature is low enough to condensate the volatile into its solid phase (ice). Around the water ice line, the temperature is high enough (>$170$\kel) that the surface energy of water ice is sufficient to allow icy particles to grow via coagulation \citep{gundlach_stickiness_2014, musiolik_contacts_2019}. Furthermore, ice sublimation at the ice line causes vapor diffusion and re-deposition outside the ice line, enhancing pebble growth and potentially triggering streaming instability \citep{drazkowska_planetesimal_2017, ros_effect_2019}. Understanding the physics and chemistry of icy pebbles and the phase changes of volatiles is thus paramount for understanding planet formation processes close to the ice lines. 

What is the composition of an icy pebble? We can retrieve some information about pebbles from comets, which are thought to be thermally unaltered objects and preserve pristine materials from the disc in their nuclei. Measurements by the \textit{Rosetta} instrument GIADA of the dust ejected from comet 67P/Churyumov–Gerasimenko (67P hereafter) constrain the porosity of the putative pebbles composing the nucleus of the comet to $\sim50$\per\,\citep{fulle_comet_2016}. Experiments on dust aggregates showed that a highly porous aggregate ($85$\per\,porosity) can be compacted down to $60$\per\,porosity, simulating compaction by collisions between pebbles in an accretion scenario \citep{weidling_physics_2009, JUTZI2009802}. Furthermore, \citet{blum_evidence_2017} suggest that pebbles with a size range of $3-6$\, mm are present in the nucleus of 67P and that the interstitial space between pebbles is filled with fluffy fractal aggregates. The measurements of the CONSERT instrument allowed deriving a dust-to-ice volume ratio of 67P in the range of $0.4-2.6$ \citep{kofman_properties_2015}. These data depict a pebble formed outside the ice line as a mm-size porous aggregate, formed by ice and small dust grains, with a variable ice-to-dust ratio.

Although pebbles are porous aggregates of refractory and volatile materials, several studies of dust evolution in protoplanetary discs handle them more simply as compact spheres made with a specific composition of volatiles and refractory compounds \citep{barriere-fouchet_dust_2005,gonzalez_accumulation_2015, drazkowska_planetesimal_2017, schoonenberg_planetesimal_2017}. This assumption may be insufficient to correctly describe coagulation processes and pebble-to-pebble collisions, but it offers a great simplicity for the purpose of modeling other dynamical processes.

The fate of icy pebbles undergoing sublimation is important to understand if the refractory materials embedded in the pebble are released back to the disc or if the sublimated pebble manages to maintain its integrity without disrupting. Many theoretical studies assume that pebbles disrupt through ice sublimation with important consequences on planetesimals growth. \citet{saito_planetesimal_2011} and \citet{ida_radial_2016} showed how disruption of icy pebbles drifting toward the central star through the ice line enhances the dust local density and triggers gravitational instability in the vicinity of the ice line. \citet{schoonenberg_planetesimal_2017} use a simple model of pebbles in their simulations, and find that disruption of pebbles through the ice line and ice deposition enhance pebble growth, in particular outside the ice line. Their pebbles are compact spheres made of $50$\per\, water ice and $50$\per\, silicates. In their work, they consider two mixtures of dust and water ice: one type of pebbles has a compact silicate core and a shell of ice (`single-seed model'), while the other one is a homogeneous mixture of \um-sized ice and silicates particles (`many-seeds model'). They assume two different behaviors of these pebbles once they drift through the ice line. The single-seed model sublimates the ice shell, leaving behind the compact silicate nucleus as a remnant. The many-seeds model sublimates the ice particles and releases small silicate particles. In the case of the many-seeds model, \citet{schoonenberg_planetesimal_2017} found an enhancement of dust-to-gas ratio interior to the ice line (but less than \citet{saito_planetesimal_2011} and \citet{ida_radial_2016}) and outside the ice line, due to the fact that the released small silicate particles couple with the gas and diffuse outward, sticking to the icy pebbles outside the ice line. Interestingly, the results from \citet{schoonenberg_planetesimal_2017} do not change considering the porosity of pebbles: increasing porosity would increase the sublimation rate, but even increasing it by a factor $100$ would not affect the consequent dust enhancement. Similarly, \citet{hyodo_planetesimal_2021} found different dust enhancement and planet formation pathways around the ice line, through pile-up of icy pebbles outside the ice line and of small silicate grains inside the ice line. The general assumption in these works is that sublimation of ice is breaking the adhesion forces between dust grains inside the pebble, causing its disruption. \citet{aumatell_breaking_2011} created in the laboratory aggregates of condensed ice crystals and studied their disruption by sublimation while levitating. They found that those fluffy aggregates are easily disrupted though a fast sublimation, but they did not include dust grains in their experimental models. Understanding if icy pebbles are disrupting throughout sublimation is essential to investigate the processes that would lead quickly to the formation of planetary embryos in the proximity of ice lines.

In addition, pebble accretion in general depends strongly on their Stokes number, and therefore on their size and structure (note that a fluffy structure could lead to a drag coefficient very different from the usually assumed drag coefficient of a sphere). Understanding the disruption of pebbles at the ice line could therefore have important consequences on the formation timescale of all planets located inside the ice line. For example, \citet{morbidelli_great_2015} explained the `great dichotomy' of the Solar System (the fact that inner planets are dry and small whereas outer planets are wet and large) by the difference between the accretion rate inside the ice line and outside it, itself resulting from both the change of mass (reduction of 50\per\,of the mass flux due to the evaporation) and of radius. Such a scenario would be put into question, at least partially, in the case where pebbles would keep their structure while crossing the ice line.

We present here new experiments on the structure of pebbles composed of water ice and silicates when they are free to sublimate in vacuum with increasing temperature. The goal of the present study is to characterize the different outcomes of sublimated icy pebbles (disruption or preservation), and understand how the environment and the composition of the pebbles determine the final outcomes of the sublimation. 

The experiments aim at answering these scientific questions:
\begin{enumerate}
 \item Are icy pebbles always disrupting due to sublimation of ice?
 \item Does the dust-to-ice ratio influence pebble disruption?
 \item What is the role of the dust size range in the disruption due to sublimation?
\end{enumerate}

The structure of the manuscript is the following. Section~\ref{sec:Methods} presents the material used, the new pebble production methods and their features, and the experimental setup used. In section ~\ref{sec:results}, we summarize the experimental results. Discussion of the results and future experiments are presented in section~\ref{sec:discussion}. Finally, we conclude with our main findings in section~\ref{sec:conclusion}.

\section{Material and Methods} \label{sec:Methods}

 The necessity of simulating cold planetary and space environments lead to the development of the SCITEAS-2 (Simulation Chamber for Imaging the Temporal Evolution of Analogue Samples version $2.0$) vacuum chamber, which provides a low-pressure and low-temperature environment for the sublimation of icy samples, with the possibility to acquire spectral measurements of the samples over time \citep{pommerol_sciteas_2015}. Two types of icy particles have been developed as analogs for icy planetary surfaces: flash-cooling of small water droplets produces two different spherical particles with size distributions peaking at $4$ and $70$\um. These particles can then be mixed with dry dust in two ways \citep{pommerol_experimenting_2019-1, poch_sublimation_2016}: through intra-mixture (i.e. dust particles embedded in icy particles) and inter-mixture (i.e. dust particles mixed externally with the icy particles). From such ice particles, it is difficult to create mm-sized aggregates. Therefore, we developed new methods for creating compact pebble-size mixtures of ice and dust.

The evolution of realistic pebbles at the ice line is a complex problem. The chemical composition of the dust, how it is mixed with ice, the dust size distribution, the presence of organics and salts, and the overall amount of water ice in the pebble are concurring to the outcome of the sublimation process.

\subsection{Dust composition} \label{sec:Dust}

Infrared observations of protoplanetary discs have shown an abundance of silicates, both in crystalline and amorphous form. In particular, Mg-rich endmembers of olivine and pyroxene are common \citep{natta_dust_2006}. Alongside silicates, materials that are more refractory are produced close to the star, such as corundum, anorthite, hibonite, perovskite, and spinel. These are the first minerals formed in the protoplanetary disc, and a certain amount can diffuse in the outer disc through processes such as turbulence, meridional flows, and X-winds \citep{watson_crystalline_2009}. The outward diffusion of a certain amount of highly refractory minerals is demonstrated by their presence in comets which never left the cold outer disc since their formation \citep{simon_refractory_2008}. 

A precise `recipe' for pebble dust mineralogy is difficult to establish, since we do not know the relative abundances of the different minerals and since they are not evenly distributed within the disc and undergo different processes during the disc lifespan. In order to compare our results to previous theoretical studies of protoplanetary disc evolution, we used two different silicates as analogs for the mineral dust: olivine and pyroxene. The list of powders used in the present work is provided in Table~\ref{tab:dust}. Different grain size ranges have been obtained by grinding the dust in a mortar and dry sieving it.

\begin{figure}
	\includegraphics[width=\columnwidth]{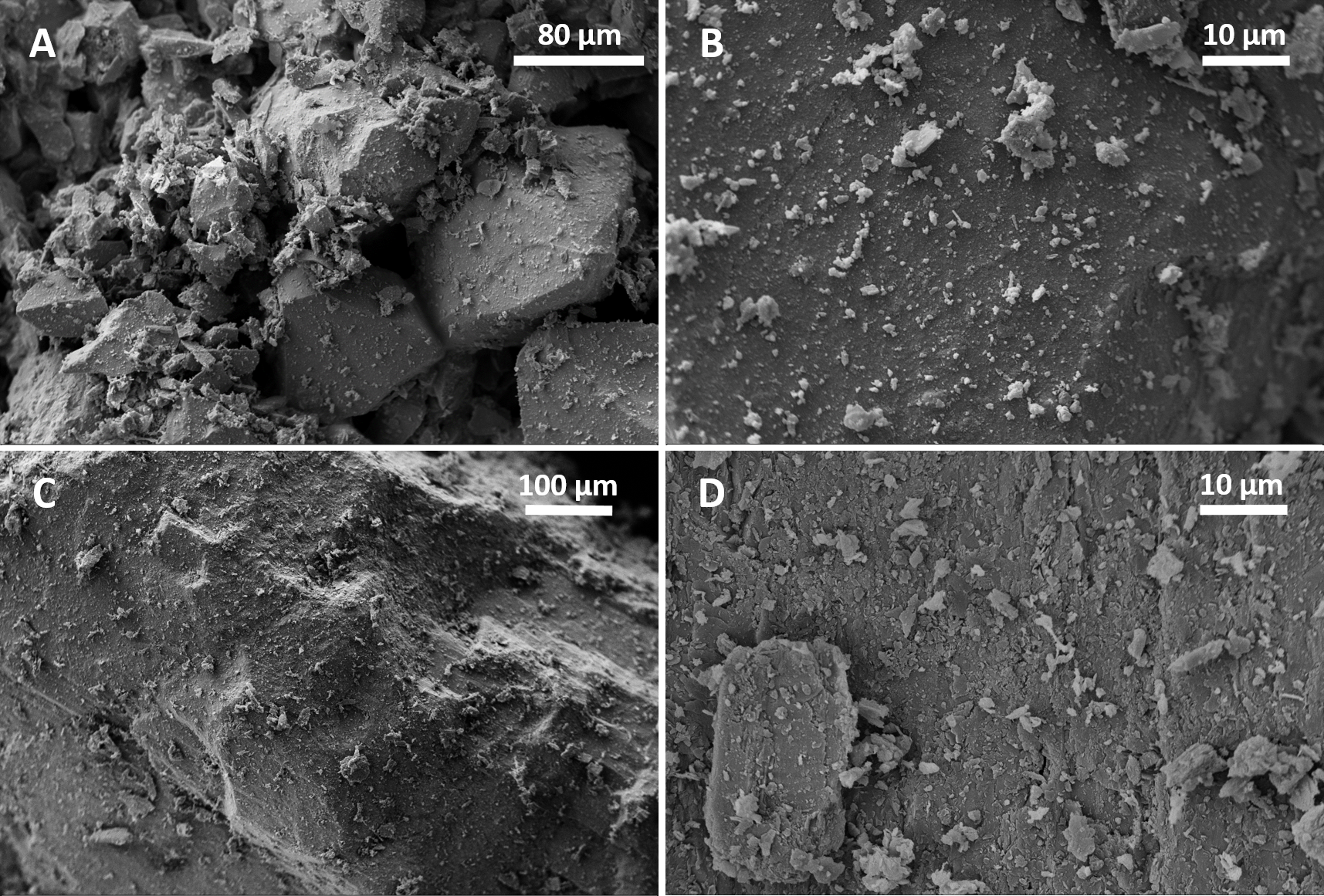}
    \caption{SEM images of olivine and pyroxene grains. A) and B) olivine dust at different magnifications. C) and D) pyroxene dust at different magnifications.}
    \label{fig:SEM}
\end{figure}

\begin{figure}
	\includegraphics[width=\columnwidth]{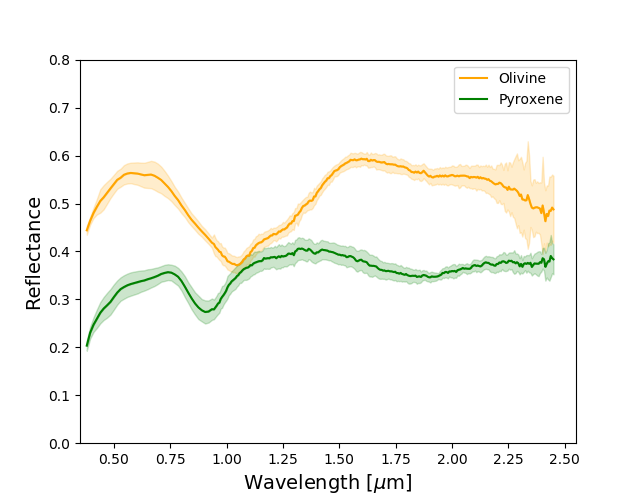}
    \caption{Visible and near-infrared reflectance spectra of olivine and pyroxene powders collected from the desiccated pebbles at the end of the sublimation experiments. The spectra do not show the absorption features associated with aqueous alteration at 1.4 and 1.9 µm. The shadowed region indicates the pixel variance of the reflectance in the selected region of interest defined to derive an average spectrum from hyperspectral cubes.}
    \label{fig:spectra}
\end{figure}

The two dust species have been analyzed through scanning electron microscopy (SEM) to check qualitatively the difference in the grain surfaces and the amount of fine material of the two silicates (Fig.~\ref{fig:SEM}). We do not see evidence for significant differences in the amount of very fine grains (<$5$ \um) and the surfaces of the grains appear relatively similar. The only notable difference is the sharper aspect of the edges of the olivine grains with respect to pyroxene. Unfortunately, it is not possible to derive the overall dust size distributions from the SEM images, which could explain some differences observed between the behaviours of pebbles made of olivine and pyroxene (see section \ref{sec:grain_size_mineralogy}). Following the sublimation experiments, we collected the desiccated dust and measured its near-infrared reflectance spectrum to look for possible signs of aqueous alteration (Fig. \ref{fig:spectra}). None of the absorption features indicative of hydration and hydroxilation (around 1.4 and 1.9 µm, see \citep{NOEDOBREA20031,Pommerol_2008}) are visible.

% table1
\begin{table}
	\centering
	\caption{List of silicates used in the experiments.}
	\label{tab:dust}
	\begin{tabular}{lcc} % four columns, alignment for each
		\hline
		Properties & Olivine & Pyroxene\\
		\hline
		Aspect & Green powder & Gray powder\\
		Initial grain size [\um] & $0-3000$ & $0-1000$\\
		Density [g\,cm$^{-3}$] & $3.3$ & $3-4$\\
		Quarry location & Norway & Brazil\\
		Provider & Microbeads AG & \begin{tabular}[c]{@{}c@{}}UCF \& Deep\\ Space Industries\end{tabular}\\
		
		\hline
	\end{tabular}
\end{table}

\subsection{Sample preparation and analysis} \label{sec:SpA}
In the rest of the paper, we refer to the two pebble models as pebble type-A (PA) and pebble type-B (PB). The main difference between the two production methods is the amount of water ice of the pebble and the relative ice-dust distribution inside the pebble. Both PA and PB are compact aggregates and their initial porosity is negligible. Table~\ref{tab:pebbles} collects the main features of PA and PB pebbles.

The two types of pebbles were analysed with X-ray computed tomography (XCT) technique, to detect the relative distribution of dust and ice inside the pebbles. The measurements have been carried on a Diondo 2 system manufactured by Diondo (Hattingen, Germany) at the Lucerne Computed tomography Imaging facility at Lucerne University of Applied Sciences and Arts. For both measurements, the microfocus tube XWT$-225$ TCHE+ from X-ray WorX has been set on an acceleration voltage of $120$\,kV and a tube current of $80$\,µA and an aluminum filter with $1$\,mm thickness was employed. The detector VAREX $4343$CT was configured with a capacity of $8$\,pF, and integration time of $1$\,s. The field of view of the detector was adjusted to match the extension of the sample with source object distance of $37$\,mm, and a source-detector distance of $851$\,mm. For the measurement of the olivine and the pyroxene sample, $1900$ and $1600$ projections of the object spinning around a fixed axis in equiangular steps were recorded, respectively. The recorded projections were reconstructed using the Feldkamp-Davis-Kress algorithm as integrated in the CERA suite from Siemens (Erlangen, Germany). The virtual cross-sections had a voxel size of $6.5$\um\, isotropically and were analysed with Volume Graphics Studio Max $3.3$ from Volume Graphics (Heidelberg, Germany) and imageJ \citep{schneider_nih_2012}.

% table2
\begin{table}
	\centering
	\caption{Summary of pebble type-A (PA) and pebble type-B (PB) main features.}
	\label{tab:pebbles}
	\begin{tabular}{lcc} % four columns, alignment for each
		\hline
		 & PA & PB\\
		\hline
		Shape & Spheroidal & Irregular spheroid\\
		Ice mass content [\per] & $50\pm10$ & $15\pm7$\\
		Typical size [mm] & $1-5$ & $2-7$\\
		Dust distribution & Homogeneous or & Homogeneous\\
		inside the pebble & in one half of the pebble & \\
		\hline
	\end{tabular}
\end{table}

\subsubsection{PA production} \label{sec:PA}
\begin{figure}
	\includegraphics[width=\columnwidth]{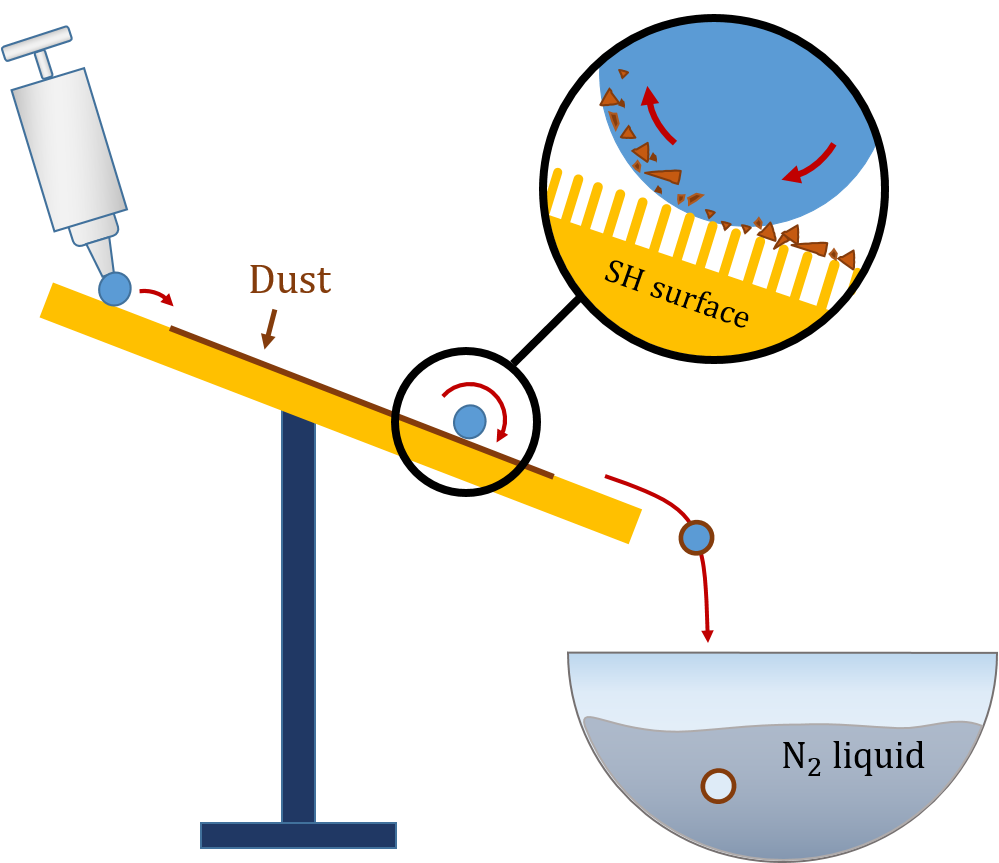}
    \caption{The procedure for PA production. A droplet created with a micropipette rolls over a superhydrophobic (SH) inclined plane, and it catches the dust deposited on it. Then, the droplet sinks into liquid nitrogen (N$_{2}$ liquid), where water freezes in a few seconds.}
    \label{fig:PA_production}
\end{figure}

The PA production method takes advantage of the behavior of water on superhydrophobic surfaces. Superhydrophobic surfaces have micrometric `spikes' which, combined with the surface tension of water, allow the liquid to rest on the top of the asperities. The water flowing on the superhydrophobic surface captures any dust deposited on the top of the spikes in a process called `self-cleaning' \citep{marmur_lotus_2004}. An inclined plane coated with superhydrophobic paint (Cytonix WX$2100$ aerosol) is placed above a bowl full of liquid nitrogen (Fig.~\ref{fig:PA_production}). On the surface, some dust is deposited with a spoon, distributing it on the surface as homogeneously as possible. A droplet of distilled water, created with a micropipette, rolls and captures the dust deposited on the plane, and then sinks in liquid nitrogen where the water freezes. 

The droplets created with the micropipette are typically about $2-3$\,mm in diameter, but sometimes a droplet impinging the plane can split and create secondary droplets of about $1$\,mm in diameter, which capture the dust in the same way as the bigger droplets. 

\begin{figure}
	\includegraphics[width=\columnwidth]{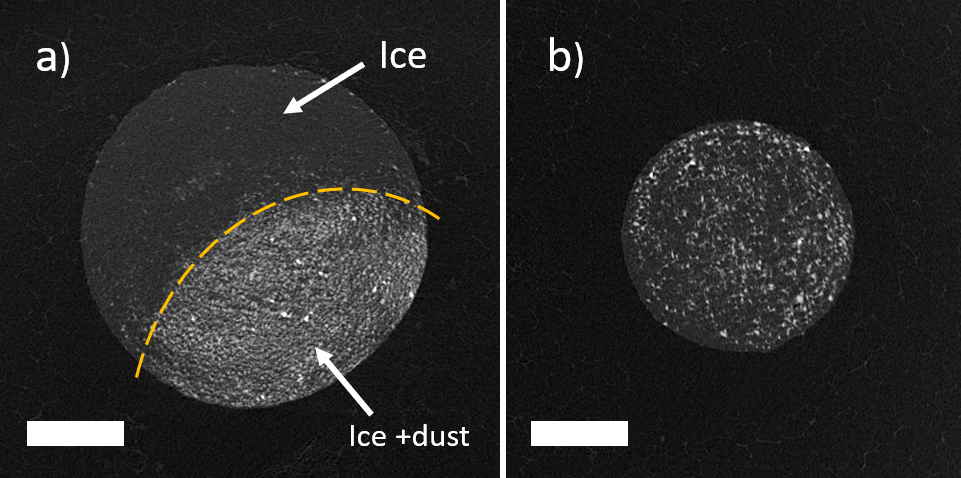}
    \caption{X-ray tomography of PA icy pebbles analogues made with olivine (dust size smaller than 50 \um). The circle in the center is the cross section of an icy pebble.  a) This PA clearly shows a heterogeneous spatial distribution of ice (dark gray material on the top) and dust (light gray material on the bottom), well distinguishable from each other. b) This PA contains an internal homogeneous distribution of dust grains. The scale bar for both images is 1 mm.}
    \label{fig:PA_XCT}
\end{figure}

The dust collected by the droplets on the plane would tend to remain on the surface of the droplet because of the centrifugal force and surface tension of water. On the other hand, if more dust is collected, the grains that are already embedded in water are gradually pushed toward the center of the droplet. In the limit case, if the droplet is saturated with dust, the water does not interact with the superhydrophobic plane anymore and the drop stops on the inclined plane. When a droplet sinks into liquid nitrogen, the inverse Leidenfrost effect \citep{thimbleby_leidenfrost_1989,adda-bedia_inverse_2016} prevents the droplet to freeze instantly, and the dust captured by the droplet can move under gravity inside the liquid core of the pebble. In this way, it may happen that the frozen pebble has a heterogeneous distribution of the dust inside it, with more dust on one half due to the gravity acting on the dust grains during the freezing process. This heterogeneity may not happen when the droplet is small (like secondary droplets): in this case, the freezing process is fast and the grains are `trapped' homogeneously inside the pebble (Fig.~\ref{fig:PA_XCT}). In this work, we refer by "heterogeneity" to the overall misplacement of grains with respect to the ice inside the pebble, while we make no assumption on the homogeneity of the spatial distribution of the grains with respect to each other.

The ice-dust ratio was estimated by weighing the pebble before and after the evaporation of the water. The ice mass was found to be in the range of $\left(50\pm10\right)$\per\, of the whole pebble mass. It can happen that a droplet does not catch much dust and has a higher ice content, but since the experiments involve several pebbles in the same sample holder, an average value is more indicative.

\subsubsection{PB production} \label{sec:PB} 
The second type of pebbles uses the capability of liquid water of linking small dust particles through capillary bridges, also called `wet granulation' \citep{iveson_nucleation_2001}. A $1$\,cm thick layer of dust is settled in a small vessel. A droplet of water is produced with a micropipette above the dust. The droplet impinges the dust layer with enough momentum to get a fast interpenetration between water and dust particles. The final wet aggregate is stable because of the capillary and viscous forces of water bridges acting on adjacent dust particles. Finally, the aggregate is carefully taken with a spoon and sunk in liquid nitrogen, where the water bridges freeze. Any dust particle that is not connected to the others through capillary bridges is lost in liquid nitrogen due to its turbulent motion at room temperature. 

\begin{figure}
	\includegraphics[width=\columnwidth]{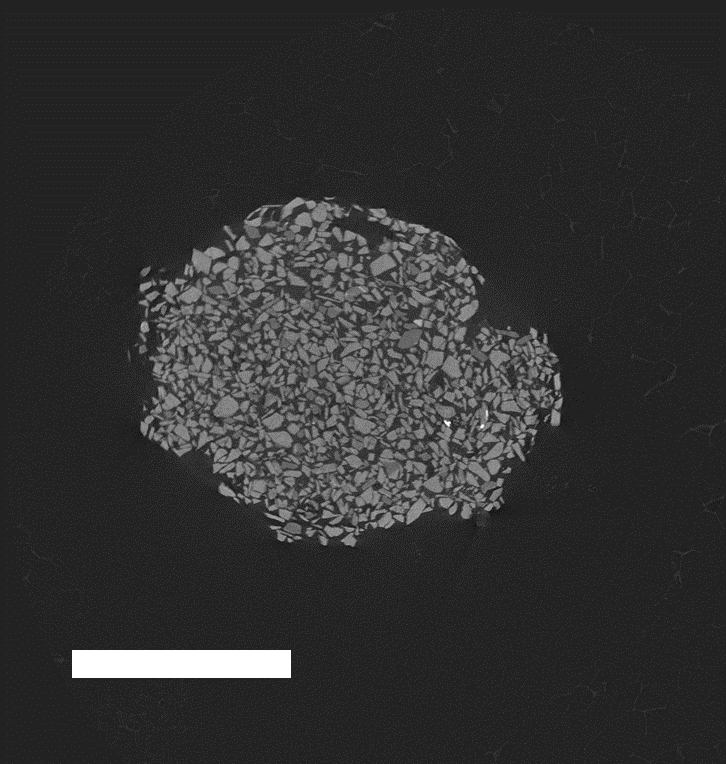}
    \caption{X-ray tomography of an icy PB made with pyroxene (dust size 100-300 \um). The icy bonds keep the dust grains (bright gray) together and the dust distribution inside the pebble is homogeneous. The scale bar is 3 mm.}
    \label{fig:PB_XCT}
\end{figure}

Weighing PBs made of different types of dust before and after the evaporation of the water resulted in an ice mass equivalent to $\left(15\pm7\right)$\per\, of the total mass. The final shape of a PB is generally an irregular spheroid with a homogeneous distribution of dust inside the pebble (Fig.~\ref{fig:PB_XCT}). Sometimes a wet aggregate disrupts into two or three pieces before being put into liquid nitrogen, resulting in a wider distribution of sizes of the final icy aggregates, ranging from $2$ to $7$\,mm in diameter. 

Granulation processes are intensively studied for industrial processes. The above procedure creates a compact pebble, where water is creating films on the surface of the particles and is filling the voids between particles \citep{iveson_nucleation_2001}. 

\subsection{Experimental setup} \label{sec:setup}
\begin{figure}
	\includegraphics[width=\columnwidth]{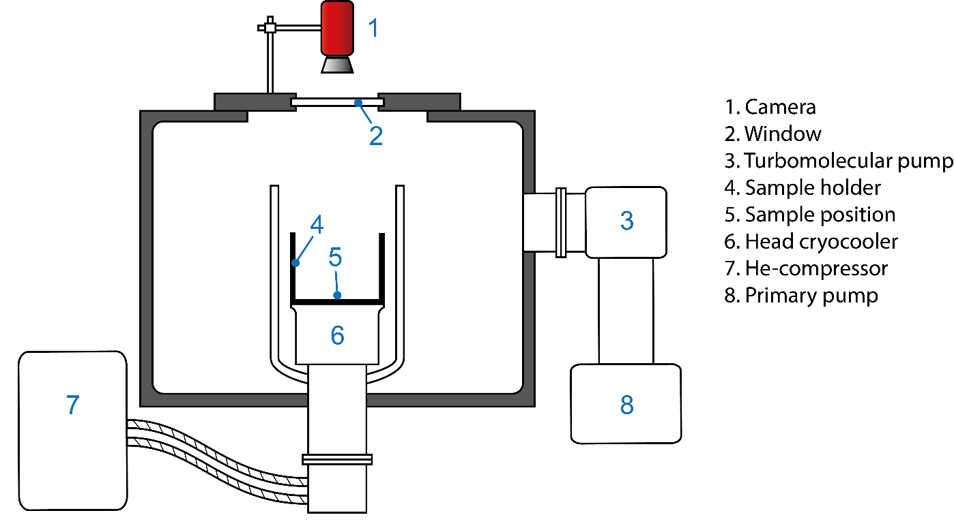}
    \caption{Scheme of the SCITEAS-$2$ setup. A camera is placed in front of a window above the vacuum chamber. The chamber is evacuated through a primary pump and a turbomolecular pump, and the sample holder is in contact with a copper head cooled down with a He-cryocooler.}
    \label{fig:SCITEAS_2}
\end{figure}

We performed all the sublimation experiments with the SCITEAS-$2$ (Simulation Chamber for Imaging the Temporal Evolution of Analogue Samples version $2.0$) vacuum chamber (Fig.~\ref{fig:SCITEAS_2}). We evacuated the chamber through a primary dry vacuum pump and a turbomolecular pump, which allowed to reach a high vacuum ($\sim10^{-7}$\mbar). At the center of the chamber, we cooled a copper plate with a Helium cryocooler, reaching temperatures down to $40$\kel. The cylindrical sample holder was placed on the copper plate and a thin sheet of graphene between the two components guaranteed good thermal contact. The icy pebbles were in contact with the sample holder, that had a higher temperature than the cooled copper plate (around $100$\kel\, at the beginning of the decompression). A PT$100$ thermal sensor measured the temperature at the bottom of sample holder, recording the temperature of the surface below the icy pebbles.

During the sample production, we pre-cooled down the copper base inside SCITEAS-$2$ by pouring liquid nitrogen into the sample holder. Once the samples were ready, we kept them in liquid nitrogen to avoid frost formation. When the copper plate of the He cryocooler was almost at the same temperature as the liquid nitrogen, we placed the samples in the cold sample holder, which was inserted in the SCITEAS-$2$ chamber. Then we plugged the PT100 sensor inside the chamber and closed the upper cover. Immediately afterwards, we turned on the primary pump to quickly evacuate the chamber, and the temperature-pressure reading started. At a pressure of about $10^{-2}$\mbar, we turned on the turbomolecular pump, allowing to reach about $10^{-6}$\mbar\, in a few minutes. The samples rested a few hours in these conditions until we achieved an equilibrium pressure and temperature inside the chamber. We then turned off the cryocooler and the samples slowly heated up by the thermal radiation coming from the window above and the walls of the chamber. The temperature increased in about $25$ hours to room temperature. At this point, we turned off the pumps to ventilate slowly the chamber. For different types of pebbles, more experiments have been carried out. 

The temperature profile of the sample holder during the experiment evolves logarithmically over time, from a temperature of around $100$\kel\, to a final temperature of about $280$\kel. The total pressure inside the chamber settles down to $10^{-7}$\mbar\, before turning off the He-cryocooler. When the temperature starts to increase, the total pressure inside the chamber increases slowly up to $\sim10^{-3}$\mbar\, due to the sublimation of ice condensed on the cold copper plate at the closure of the chamber. At around $10$ hours after the beginning of the experiments, it slowly decreases down to $\sim10^{-6}$\mbar.

\subsubsection{Disruption measurement} \label{sec:disruption_measurement} 
\begin{figure}

	\includegraphics[width=\columnwidth]{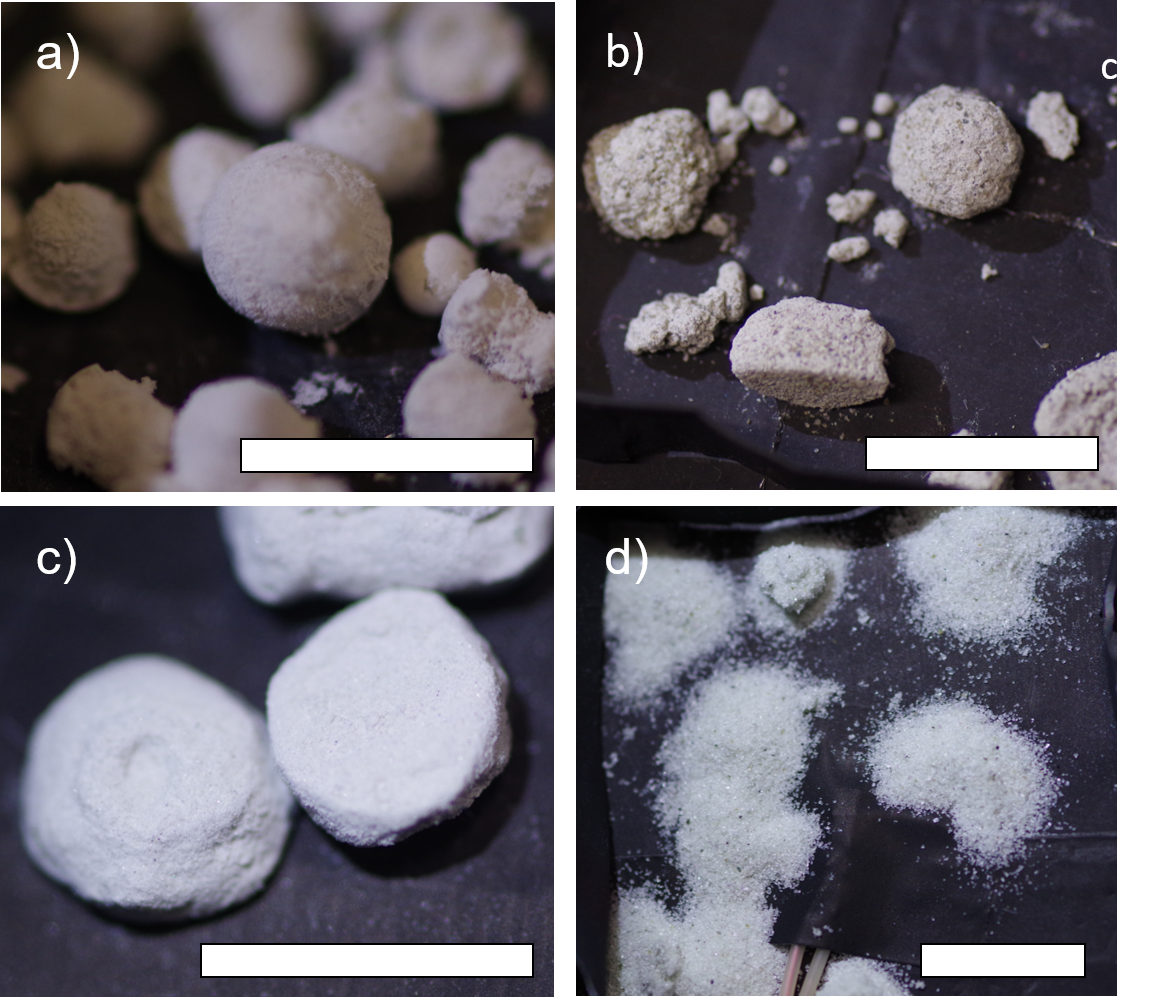}
    \caption{Different outcomes of the experiment: pebbles can preserve their 3D shape or disrupt in piles of dust. a) Preserved PA made of olivine with grain size range < 50 \um. b) Preserved PB made with pyroxene. c) preserved PB made with olivine. d) Disrupted PB made with coarse dust of pyroxene. Note a partially disrupted pebble on the top of the image, surrounded by a halo of settled dust whit a bigger projected area than the original pebble. All scale bars are adjusted to be about 1 cm.}
    \label{fig:pebbles_pictures}
\end{figure}

After the sublimation process inside the chamber, pebbles can either keep their spheroid shape or disrupt completely or partially in piles of dust. Both gravity and sublimation participate to this disruption process (see section ~\ref{sec:gravity}). In Fig.~\ref{fig:pebbles_pictures} we show how pebbles look like once removed from the chamber when the experiment is finished. The preserved pebbles show a clear spheroid shape conserving overhangs and protrusions, while the disrupted pebbles are just piles of dust and lost their initial spheroidal shape.

Analysing the pebbles outside the chamber is difficult.The vibrations caused while removing the sample holder from the vacuum chamber can disrupt weak preserved pebbles, and the re-compression of the chamber after the experiment could as well participate in the disruption of weakly preserved dust aggregates. To address this problem, we developed a method for evaluating the disruption of the aggregates directly inside the chamber, without handling the samples. To determine quantitatively the level of preservation or disruption of the pebbles, we placed a camera above the window, pointing at the sample. Two images were taken at the beginning and at the end of the experiment, allowing the detection of differences in the projected area of the pebbles before ($A_{\mathrm{b}}$) and after ($A_{\mathrm{a}}$) sublimation. For geometrical reasons, when a spheroidal aggregate disrupts into a pile of dust, its projected area increases. We define then the disruption parameter $D$ as follows:
\begin{equation}
    D=\frac{A_{\mathrm{b}}}{A_{\mathrm{a}}}.
	\label{eq:D}
\end{equation}

\begin{figure}
	\includegraphics[width=\columnwidth]{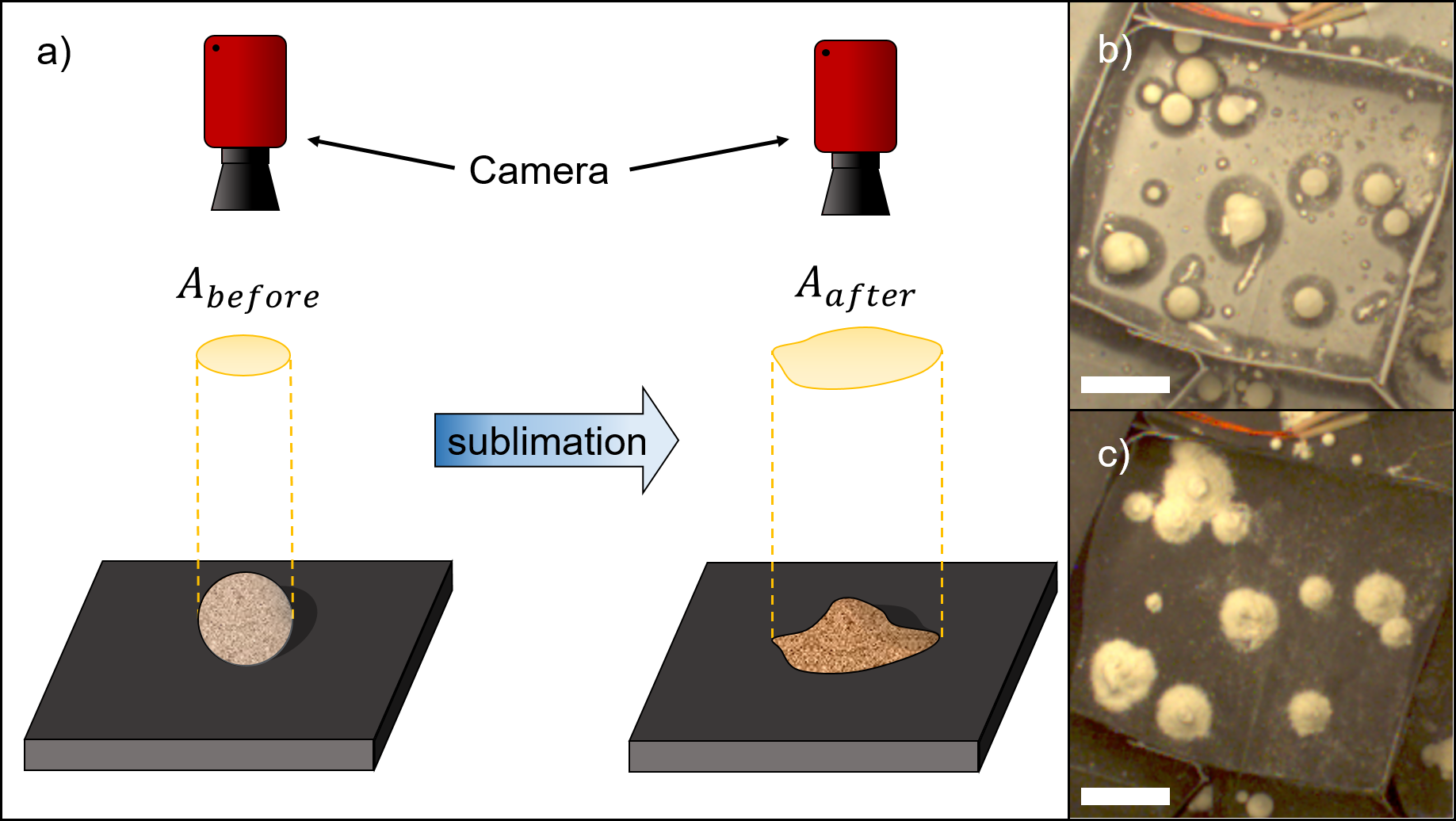}
    \caption{a) Setup for observation of pebble disruption. The ratio between $A_{\mathrm{b}}$ and $A_{\mathrm{a}}$ provides information on the level of disruption of the pebble. The images before and after sublimation are used to calculate the ratio between projected areas. b) PAs of olivine before sublimation. c) The same pebbles after sublimation. The disruption created piles of dust with bigger cross-sectional area. The scale bar is 1 cm.}
    \label{fig:disruption}
\end{figure}

If the ratio of the areas is close to $1$, then the pebble shape did not change and the pebble survived the sublimation of the ice. If the ratio is $<1$ then the projected area increased and the pebble disrupted in a pile of dust (Fig.~\ref{fig:disruption}). Disruption is measured at the end of the sublimation, before the re-compression of the chamber and its ventilation. This method allows to recognize small changes in the projected areas of disrupted pebbles, leading to a detection of disruption. For each type of pebbles and dust combination, we have run sufficient sublimation experiments to obtain a sample with good statistic relevance.
The pebbles that do not disrupt during the experiment, can resist months at room temperature and ambient pressure, until some external forces stronger than gravity break the aggregates (handling, vibrations).

\section{Results}\label{sec:results}
The sublimation outcome of icy pebbles depends on the grain size ranges of the dust, the ice content (i.e. PA or PB), the silicate type, and the dust-ice distribution inside the pebbles. The experiments aim at distinguishing the effects of each of these contributions to the final preservation or disruption of the pebble after sublimation. 

The disruption parameter $D$ is averaged between pebbles of the same type (PA or PB, dust type, and grain size range). The maximum error related to the average $\bar{D}$ is:
\begin{equation}
    \Delta \bar{D}=\sqrt{{\sigma_{P}}^{2}+4\per \bar{D}}
	\label{eq:errD}
\end{equation}
where $\sigma_{P}$ is the standard deviation of $D$ between pebbles of the same type, and, assuming $2\per$\, relative error on the experimental evaluation of the pebble area on the image, the term $4\per \bar{D}$ is the absolute error of the average disruption parameter measured for pebbles of the same type. When the measurement of $D$ with its maximum error is consistent with $D=1$, then the pebbles are not disrupting. 

\subsection{Pebbles of olivine and pyroxene with different grain size ranges}
\label{PA_PB_o_p}
\begin{figure}
	\includegraphics[width=\columnwidth]{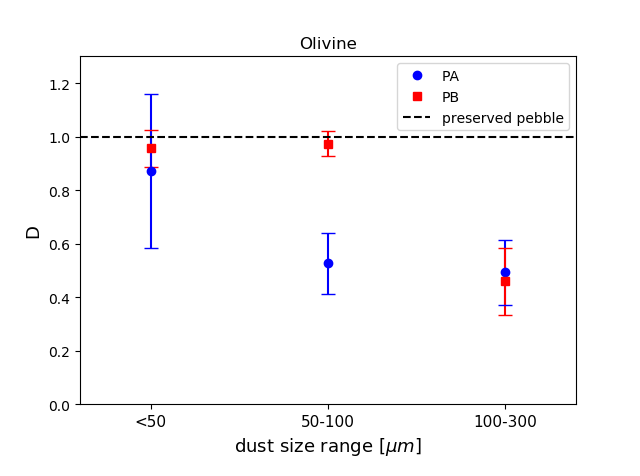}
    \caption{Disruption parameter depending on grain size range of olivine, for both PA and PB. When the measurement of $D$ is consistent with $1$, then the pebbles survived the sublimation process.}
    \label{fig:olivine}
\end{figure}

\begin{figure}
	\includegraphics[width=\columnwidth]{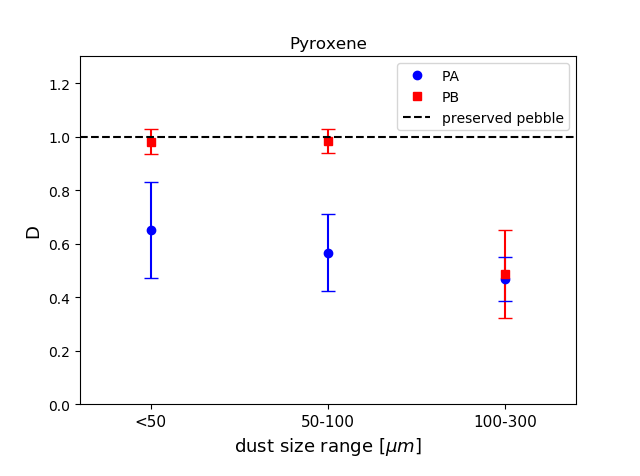}
    \caption{Disruption parameter depending on grain size range of pyroxene, for both PA and PB. When the measurement of $D$ is consistent with $1$, then the pebbles survived the sublimation process.}
    \label{fig:pyroxene}
\end{figure}

We created the two types of pebble (PA or PB) with different grain sizes for the two dust types (olivine and pyroxene). The grain size ranges are obtained by grinding the powder in a mortar and dry sieving the dust with nylon sieves. Three ranges are used in the present work: grains smaller than $50$\um\, (referred to as `fine component'), $50-100$\um, and $100-300$\um\, (referred to as `coarse component'). Olivine pebbles made with dust grains smaller than $50$\um\, are not affected by sublimation (Fig.~\ref{fig:olivine}). At $50-100$\um, we already see that PAs do not maintain their integrity, while PBs are still preserved. For coarser dust, $100-300$\um, both PAs and PBs are not able to counter ice sublimation and gravity and they disrupt. Pyroxene pebbles (Fig.~\ref{fig:pyroxene}), on the other hand, show complete disruption for PAs, for all the size ranges. PBs however are following the same pattern as olivine pebbles: the pebbles survive up to $100$\um, and the coarse component is disrupting under sublimation pressure and gravity.

Big error bars are usually associated to the presence, in the same sample, of both disrupted and preserved pebble (e.g. PA with olivine). This is leading to a bigger standard deviation of $D$ within the sample. 

\subsection{Pebbles of olivine and pyroxene with two-component dust size ranges}
\begin{figure}
	\includegraphics[width=\columnwidth]{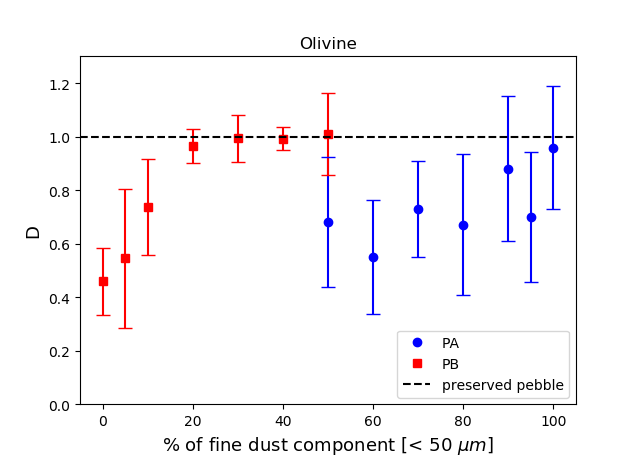}
    \caption{Disruption parameter depending on the amount of fine component ($<50$\um) mixed with coarse component ($100-300$\um) of olivine for both PA and PB. When the measurement of $D$ is consistent with $1$, the pebbles survived the sublimation process.}
    \label{fig:olivine_dist}
\end{figure}

\begin{figure}
	\includegraphics[width=\columnwidth]{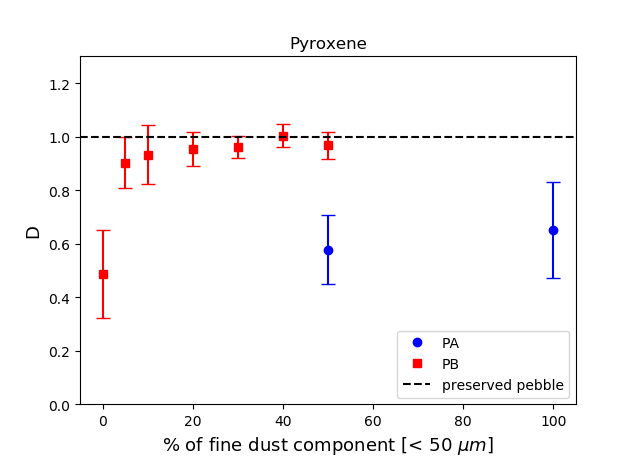}
    \caption{Disruption parameter depending on the amount of fine component ($<50$\um) mixed with coarse component ($100-300$\um) of pyroxene for both PA and PB. When the measurement of $D$ is consistent with $1$, the pebbles survived the sublimation process.}
    \label{fig:pyroxene_dist}
\end{figure}

Disruption of pebbles with $100-300$\um\, dust grains of olivine and pyroxene is observed for both PA and PB. Mixing a certain amount of the fine component ($<50$\um) with the coarse component ($100-300$\um) can change this outcome drastically (Fig.~\ref{fig:olivine_dist} -~\ref{fig:pyroxene_dist}):
\begin{enumerate}
\item PBs made with olivine survive if $20$\per\, of the dust mass is the fine component;
\item PAs made with pyroxene are always disrupting, even if they are made with $100$\per\, of fine dust (see also Fig.~\ref{fig:pyroxene});
\item PBs made with pyroxene do not disrupt through sublimation when at least $5- 10$\per\, mass of dust is $<50$\um;
\item PAs made with olivine disrupt unless the fine dust component is up to $90- 100$\per\, of the total dust mass.

\end{enumerate}

This is an indication of the capability of fine dust to cement bigger particles together after sublimation. If the icy pebbles contain a higher amount of ice (PAs), then the amount of fine dust needed to cement bigger grains increases or in extreme case sublimation disrupts the pebbles even if they are made of small grains only.
\section{Discussion} \label{sec:discussion}
In this work, we present sublimation experiments with two different icy pebble models to investigate how the initial amount of ice in the pebble, the dust type, and the dust size ranges are affecting the disruption of the icy pebble through the sublimation process.
In the case of an intimate mixture of ice and dust, icy aggregates are thought to be unstable upon sublimation of the ice and undergo disruption releasing small silicate particles \citep{saito_planetesimal_2011, schoonenberg_planetesimal_2017, hyodo_planetesimal_2021}. Our experiments show that it is possible to maintain the integrity of pebbles after the complete sublimation of the ice intermixed between the dust grains.
\subsection{Sublimation rate} \label{sec:sublimation}
Inside the vacuum chamber SCITEAS-$2$, the temperature below the pebbles rises from $\sim120$\kel\, to $280$\kel\, in about $20$ hours. The sublimation of the ice occurs during this time, although the experimental setup does not allow detecting exactly when it ends. 

The mean size of PAs is about $1.5$\,mm in radius, with an ice volume ratio of $77$\per. The equivalent ice would be contained into an equivalent sphere with a radius of around $1.4$\,mm. 
The linear decrease rate of ice in the direction of the surface normal $R_{\mathrm{ice}}$, can be calculated with the Hertz-Knudsen formula \citep{knudsen_gesetze_1909},
\begin{equation}
    R_{\mathrm{ice}}=\frac{V_{\mathrm{m}}(p-p_{\mathrm{sat}})}{\sqrt{2\pi\mu R_{\mathrm{g}}T}}
	\label{eq:knudsen}
\end{equation}
where $V_{\mathrm{m}}$ is the molar volume, $p$ is the vapor pressure above the surface, $p_{\mathrm{sat}}$ the saturation pressure, $\mu$ is the molar mass of ice, $R_{\mathrm{g}}$ is the gas universal constant, and $T$ is the temperature of the ice. The saturation pressure $p_{\mathrm{sat}}$ is related exponentially to the temperature $T$ \citep{haynes_condensation_1992}
\begin{equation}
    p_{\mathrm{sat}}=6.034\cdot10^{12}\mathrm{g\,cm^{-1} s^{-2}}\cdot\mathrm{e}^{-\frac{5938\mathrm{\kel}}{T}}.
	\label{eq:sat_p}
\end{equation}
The total pressure inside the vacuum chamber varies over time (from $10^{-7}$\mbar\, to $10^{-3}$\mbar), depending on the sublimation processes of the condensed gasses trapped inside. Since at the beginning of the experiment some vapor condenses inside the chamber on the cold copper plate, when the temperature increases during the experiment, the pressure increases as well, until all the condensed ice is sublimated. Assuming that the pressure inside the chamber is due to the sublimation of such ice, the vapor partial pressure $p$ is exactly the pressure measured during the experiment inside the vacuum chamber. Once the partial vapor pressure is known, the rate of linear decrease of ice can be calculated as a function of the temperature only. Assuming that the ice of the pebbles is at the same temperature of the sample holder on which they lay and since the temperature of the sample holder is known, it is possible to calculate at what temperature the sublimation happens and how much time it requires. For our experiments, the sublimation time of a compact ice sphere with a radius of around 1.4 mm is between 4 and 6 hours and it happens on average when the temperature spans from $\sim190$ to $240$\kel. This gives a sublimation rate of about $2\cdot10^{-1}$ mm/hour.

At the water ice line temperature, the sublimation rate is considerably slower than in the laboratory. If we assume a protoplanetary disc with gas surface density $\Sigma(r)$ and a temperature profile $T(r)$ described by
\begin{equation}
    \Sigma(r)=1000\,\mathrm{g\,cm^{-2}}\left(\frac{r}{\mathrm{au}}\right)^{-1}
	\label{eq:sigma}
\end{equation}
\begin{equation}
    T(r)=280\mathrm\kel\left(\frac{r}{\mathrm{au}}\right)^{-\frac{1}{2}}
	\label{eq:T}
\end{equation}
where $r$ is the radial distance from the central star, and the amount of water (in both ice and vapor phases) with respect to the hydrogen and helium content of the disc is equal to $Z_\mathrm{H_{2}O}=0.005$, it is possible to calculate the partial vapor pressure $P_{v}$ from:
\begin{equation}
    P_{v}=\frac{k_\mathrm{b}Z_\mathrm{H_{2}O}T(r)\Sigma(r)}{\mu_\mathrm{H_{2}O}\sqrt{2\pi} H_\mathrm{gas}(r)}
	\label{eq:P}
\end{equation}

where $k_\mathrm{b}$ is the Boltzmann constant, $\mu_\mathrm{H_{2}O}$ is the weight of a molecule of water and $H_\mathrm{gas}(r)$ is the disc scale height given by
\begin{equation}
    H_\mathrm{gas}(r)\approx0.033\,\mathrm{au}\left(\frac{r}{\mathrm{au}}\right)^{\frac{5}{4}}
	\label{eq:H}
\end{equation}
Given these parameters, the ice line is around $2.7$\,au, and the sublimation rate at $2.5$\,au (just inside the ice line), is about $10^{-4}$mm/hour.

Considering an amount of water equal to $Z_\mathrm{H_{2}O}=0.005$ in the disk, the water vapor pressure at the ice line is within the range of the vacuum chamber pressure (around $10^{-6}$\mbar), but the temperature of the sample holder increases quickly above the ice line temperature ($\sim170$\kel) up to $280$\kel\ in 20 hours. This means that pebbles undergoing sublimation in the laboratory are at least three orders of magnitude faster in sublimating their ice than pebbles just inside the ice line, highlighting the fact that survivability of icy pebbles is possible at a much higher sublimation rate than the one found at the ice line. It is probable that some of the pebbles that are disrupting in our experiment could survive a sublimation process in conditions of slower sublimation rates and in the absence of gravity. Nonetheless, temperature fluctuations in protoplanetary discs could lead to faster sublimation rates than the one at the ice line (see e.g. \citet{mcnally_temperature_2014} and \citet{bodenan_can_2020}).

Previously, theoretical works have assumed that pebbles disrupt after sublimation freeing micrometer-size dust particles inside the ice line, leading to enhancement of planet formation just inside and outside the ice line. What would happen if a pebble similar to a PA (in size and porosity) passes the ice line without disrupting? If pebbles are not disrupting through sublimation, then planet formation due to pebble accretion could be enhanced inside the ice line. This is a direct effect of the change of the Stokes number of pebbles or dimensionless friction time of the pebble in the Epstein regime,
\begin{equation}
    St=\frac{a\rho_{\mathrm{p}}}{H_\mathrm{gas}\rho_{\mathrm{gas}}}
	\label{eq:Stokes}
\end{equation}
with $a$ particle radius, and $\rho_{\mathrm{p}}$ and $\rho_{\mathrm{gas}}$ being the pebble and gas density. When the ice sublimates, the pebble changes its apparent density, while maintaining the same size. The Stokes number of a sublimated olivine pebble with $1.5$\,mm of radius and 77\per\, porosity (see section~\ref{sec:porosity}) is about $5\cdot10^{2}$ times the Stokes number of a \um-size olivine grain. Since the effective radius for accretion onto planetesimal depends on the Stokes number of the accreting pebbles according to $r_{\mathrm{eff}}\propto St^{\frac{1}{3}}$, their accretion efficiency would be increased inside the ice line \citep{lambrechts_rapid_2012, morbidelli_great_2015}. On the other hand, a higher Stokes number means a higher drift velocity toward the star, decreasing the efficiency of planet formation from pile-up of material close to the ice line. It should be noted, however, that higher velocities of pebbles could result in disruptive pebble-pebble collisions, so pile-up of fine dust material could still be possible.

\subsection{Role of porosity} \label{sec:porosity}
Many studies show that porosity is important for pebble evolution and for hit-and-stick growth \citep{dominik_physics_1997, ormel_dust_2007, garcia_evolution_2020}. For our experiments, we use two pebble models that have negligible porosity at the beginning of the sublimation experiment. At the end of the experiment however, the porosity of the preserved pebbles increases since all the ice sublimated, but they maintain their shape with insignificant changes. We can therefore calculate the final porosity of pebbles $\Phi_{\mathrm{f}}$:
\begin{equation}
    \Phi_{\mathrm{f}}=1-\frac{V_{\mathrm{dust}}}{V_{\mathrm{pebble}}}=\frac{i}{i+(1-i)\frac{\rho_{\mathrm{ice}}}{\rho_{\mathrm{dust}}}}
	\label{eq:porosity_final}
\end{equation}
Where $V_{\mathrm{dust}}$ is the volume occupied by the dust, $V_{\mathrm{pebble}}$ is the total volume of the pebble, $\rho_{\mathrm{ice}}$ and $\rho_{\mathrm{dust}}$ are the ice and dust densities respectively, and $i$ is the ice mass percentage inside the initial pebble. The final porosity of PAs is $\left(77\pm7\right)$\per\, and the final porosity of PBs is $\left(37\pm12\right)$\per. PBs porosity is close to the expected porosity of pebbles in the nucleus of 67P \citep{fulle_comet_2016}, and their initial dust-to-ice volume ratio is $1.7\pm0.4$ which is consistent with the limits found by \citet{kofman_properties_2015}.

After sublimation, preserved pebbles are porous, highlighting that such kind of dry porous aggregates would survive sublimation with an amount of ice filling the space between dust grains up to the initial mass ratio (50 and 15\per\, mass of ice for PAs and PBs, respectively). In the limit case, the ice is filling all the space inside the dry porous pebble, but sublimation is not strong enough to push apart the dust particles. 

\subsection{Role of gravity}\label{sec:gravity}
In the laboratory, sublimation and gravity are both concurring to the disruption of pebbles. Ice sublimation pushes the grains apart and tends to disrupt the pebble, while gravity disrupts any grain-to-grain adhesion that is weaker than the gravity force of the dust column above it inside the pebble.

Hereafter, we refer to gravity as a "disrupting" force. There are many configurations of granular materials where the dust is kept in a certain position because of gravity which acts as a "constructive force". For example, a dust pile maintains its shape because of the static friction between grains, which counters actively the tangent force of gravity and prevents the grains to slide further. On one hand, if we consider a spheroidal aggregate of dust, the top hemisphere of the pebble is similar to a pile of dust, where friction and adhesion between particles are preventing them to slide and roll down. On the other hand, the bottom hemisphere can be seen as an upside-down pile of dust. Gravity and friction forces would tend to disrupt such configuration of dust in a pile of dust, and the weight of the top-half hemisphere of dust contributes to the disruption of the bottom-half hemisphere. This is not observed for preserved pebbles (see Fig.~\ref{fig:pebbles_pictures} a-b-c). The bottom-half of the pebble is well preserved and overhangs and protrusions are conserved in the final shape. Therefore, gravity has a disruptive role in transforming pebbles in piles of dust and preserved pebbles survived both gravity and sublimation forces during the experiment.

While it is clear that preserved pebbles survived both forces, disruption of pebbles in our experiments could be caused by both sublimation and gravity and there is a regime for which sublimation is more important than gravity. If we assume a spherical dust grain resting on an icy surface, we can compare the gravity force to the drag force of the sublimated ice acting on the small grain. The drag force $F_{\mathrm{d}}$ in Epstein regime is given by 
\begin{equation}
    F_{\mathrm{d}}=\frac{\pi}{3}\rho_{\mathrm{vap}}{R_{\mathrm{p}}}^{2}\sigma_{\mathrm{th}}\nu
	\label{eq:epstein_drag}
\end{equation}
with $\rho_{\mathrm{vap}}$ vapor density just above the surface of the pebble, $R_{\mathrm{p}}$ dust particle radius, $\sigma_{\mathrm{th}}$ thermal velocity, and $\nu$ flux velocity of the sublimating gas. The flux velocity can be retrieved applying the conservation of mass at the surface of the pebble passing from ice to vapor
\begin{equation}
    \mu\rho_{\mathrm{ice}}R_{\mathrm{ice}}=\rho_{\mathrm{vap}}\nu
	\label{eq:mass_conservation}
\end{equation}
where $\rho_{\mathrm{ice}}$ is the ice density (in mol\,$\mathrm{m}^{-3}$), $R_{\mathrm{ice}}$ is the sublimation velocity (see equation ~\ref{eq:knudsen}) and $\rho_{\mathrm{vap}}$ is given by the ideal gas law
\begin{equation}
    \rho_{\mathrm{vap}}=\frac{p_{\mathrm{sat}}}{RT_{\mathrm{ice}}}
	\label{eq:gas_law}
\end{equation}
with $p_{\mathrm{sat}}$ saturation pressure (see equation ~\ref{eq:sat_p}), and $T_{\mathrm{ice}}$ temperature of the ice. The thermal velocity $\sigma_{\mathrm{th}}$ is
\begin{equation}
    \sigma_{\mathrm{th}}=\sqrt{\frac{8RT_{\mathrm{ice}}}{\pi\mu_{\mathrm{H_{2}O}}}}
	\label{eq:th_velocity}
\end{equation}

\begin{figure}
	\includegraphics[width=\columnwidth]{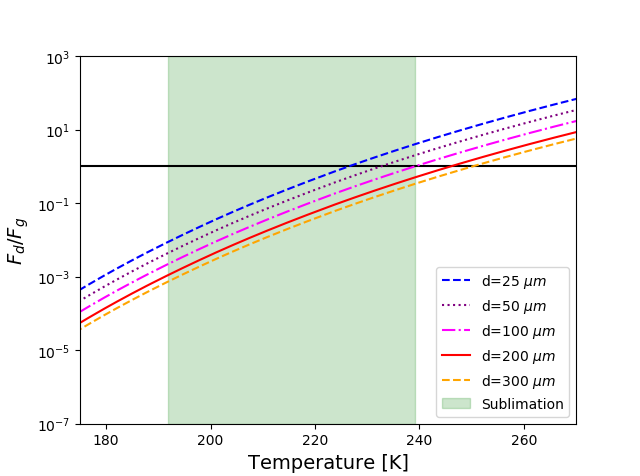}
    \caption{Ratio between sublimation drag force $F_{\mathrm{d}}$ and gravity force $F_{\mathrm{g}}$ acting on particles of different diameters vs. temperature of the ice. The sublimation of the pebble happens between $190$ and $240$\kel, and in this range of temperatures, gravity is the main disruption force for coarse grains, while sublimation is important in the higher range of temperatures for grains smaller than $100$\um. The black continuous line is a reference for the case in which sublimation drag force and gravity are equal.}
    \label{fig:drag_force}
\end{figure}

 If we assume that the temperature of the ice is the same of the sample holder, it is possible to calculate the drag force during the experiment and compare it to the gravity force acting on different grain sizes. Fig.~\ref{fig:drag_force} shows how the ratio between drag force and gravity force is varying depending on the temperature of the ice, for different grain diameters. In the temperature range where sublimation is important for the pebbles, gravity is the disrupting force in the case of coarse grains, while the drag force due to sublimation dominates the gravity force for small grains (up to $100$\um) above $\sim225$\kel. This means that all the preserved pebbles with grains smaller than $100$\um\, overcame sublimation-driven disruption. Gravity force can be still important for preserved pebbles if inside the pebble there is a heterogeneous dust-ice distribution. In this case, if the pebble survives sublimation, it can still collapse under gravity because of a structural deficiency, like a hole left by the ice inside the core of the pebble. This is why the error on the disruption parameter for PAs is large, although the general trend for different dust size ranges is not affected because of the presence of pebbles that are homogeneous and do not have any structural weaknesses when the ice is gone. 
 
A detailed description of the sublimation-driven disruption process should take into consideration what happens inside the pebble, and the possible formation of over-pressure in pores of the pebbles, but this kind of model is beyond the purpose of the present work, and will the be topic of future investigations.
 
\subsection{Role of grain size ranges and mineralogy}\label{sec:grain_size_mineralogy}
For our experiments, we used two types of silicates: olivine and pyroxene. The dust grain size ranges considered in this work span from $0$ to $300$\um, divided in three categories: ‘fine component’ (less than $50$\um), $50-100$\um, and ‘coarse component’ ($100-300$\um). In protoplanetary discs before growth to pebbles size, dust is mostly present as sub-\um\, or \um-size covering a narrower size range than our ‘fine component’ \citep{bouwman_processing_2001,boekel_grain_2003}. Nonetheless, our experiments show that pebbles can overcome sublimation with a dust size up to $50$\um\, in the case of olivine PAs and $100$\um\, in the case of PBs. 

A pebble made with the coarse component always disrupts, but when mixed with a certain amount of fine component it can be preserved through sublimation. The minimum amount of dust fine component needed for preservation varies with the mineral species, being higher for PB olivine (about $20$\per\, of the total dust mass) and lower for PB pyroxene ($10$\per\, of the total dust mass). The pyroxene PAs are always disrupting with any percentage of fine dust, while olivine PAs increase their stability through sublimation when the fine component is between $90$ and $100$\per\, of the total dust mass. Some olivine PAs do not disrupt, increasing the variability of the disruption parameter within the sample, and therefore the total error associated with the measurement. Therefore, the measurement of olivine PAs shows an increasing disruption parameter with increasing amount of fine component, but the exact minimum amount of fine component needed to avoid disruption is difficult to retrieve. In general, the introduction of fine dust mixed with coarse dust increases the pebble stability through sublimation. The small particles could act as bridges between bigger particles, increasing their adhesion capability in a way similar to the one described by \citet{seiphoori_formation_2020}. Future laboratory work should address this point, by observing the microscopic morphology of the icy pebbles and their sublimated remnants in loco.

Olivine and pyroxene pebbles have slightly different behaviors. PAs with dust size smaller than $50$\um\, are preserved for olivine and disrupted for pyroxene (Fig.~\ref{fig:olivine} -~\ref{fig:pyroxene}). Furthermore, when fine and coarse dusts are mixed, olivine PBs seem to necessitate a higher percentage of fine component than pyroxene PBs to be preserved (Fig. ~\ref{fig:olivine_dist} -~\ref{fig:pyroxene_dist}). Explaining these differences is challenging, since several factors are governing the adhesion strength of particles. The ability of grains to aggregate and to form bonds between each other is given by properties of the material (such as surface energy, Poisson ratio and Young’s modulus), grain shape characteristics (surface roughness, crystal faces) and physical and chemical state of grain surfaces (adsorption layers, chemical reactions, electric charges). Our SEM measurement do not show significant alterations of the grain surfaces before and after the experiment, and the spectra of the grains after the experiment exclude the presence of large quantity of hydroxylated material (Fig. ~\ref{fig:spectra}), which could also weaken grains bonding \citep{quadery_role_2017}. Although the SEM images do not show clear difference in the amount of very fine material (<$5$ \um), a difference in particles size distribution could explain the different behaviour of olivine and pyroxene, leading to a more adhesive behaviour of pyroxene over olivine (Fig. ~\ref{fig:olivine_dist} -~\ref{fig:pyroxene_dist}). It should be noted, however, that the adhesion capability of the two species are determined also by other parameters cited above (e.g. grain roughness).

Future experiments will address the different behavior of pyroxene and olivine, and the role of ice interaction with grains of different sizes.

\section{Conclusions} \label{sec:conclusion}
Different studies show that it is possible to enhance the growth of dust inside and outside the ice line, if icy pebbles are assumed to completely disrupt by sublimation inside the ice line \citep{saito_planetesimal_2011,schoonenberg_planetesimal_2017,hyodo_planetesimal_2021}. We researched experimentally the outcome of two icy pebble models undergoing sublimation in low-temperature and low-pressure conditions varying the dust mineralogy, the water ice content, and the dust size range. 

Our findings can be summarized as follows:

\begin{enumerate}
\item Icy pebbles can survive sublimation. We demonstrate that a range of combinations of dust type (olivine or pyroxene), dust size ranges and ice content lead to preserved pebbles upon sublimation of the ice.
\item Pebbles with low ice content survive sublimation better than the ones with higher ice content. Our results show that pebbles with 15\per\, ice mass are more resistant to disruption than pebbles with 50\per\, ice mass for different dust size ranges.
\item Icy pebbles survive better sublimation if the dust particles are smaller than $50$\um. If the pebble is made with coarse particles ($100-300$\um), a minimum amount of fine dust allows it to avoid disruption. The minimum amount of fine dust increases with the amount of ice and is dependent on the silicate type. 
\end{enumerate}

These results are relevant for planet formation processes close to the water ice line and downstream of it, providing useful information for modeling the behavior of ice-dust aggregates in protoplanetary discs when sublimation is occurring. In particular, they provide elements for modelling the sublimation outcome of icy pebbles depending on their ice content, dust type, and dust size ranges, which play a major role in the disruption process. Verifying these results for other dust types, narrower dust size distributions, and in presence of organics is an important future extension of this work. 

\section*{Acknowledgements}
This work has been carried out within the framework of the NCCR PlanetS supported by the Swiss National Science Foundation. We thank the two anonymous reviewers for the useful insights which helped improving the manuscript.

%%%%%%%%%%%%%%%%%%%%%%%%%%%%%%%%%%%%%%%%%%%%%%%%%%
\section*{Data Availability}

All the camera images of the sublimation process are available in figshare from \url{https://figshare.com/articles/figure/The_fate_of_icy_pebbles_undergoing_sublimation_in_protoplanetary_disks/16912399}.


\typeout{}
\bibliographystyle{mnras}
\bibliography{bibliography} % if your bibtex file is called example.bib

%\begin{thebibliography}{99}
%\end{thebibliography}

%%%%%%%%%%%%%%%%%%%%%%%%%%%%%%%%%%%%%%%%%%%%%%%%%%

%%%%%%%%%%%%%%%%% APPENDICES %%%%%%%%%%%%%%%%%%%%%

%%%%%%%%%%%%%%%%%%%%%%%%%%%%%%%%%%%%%%%%%%%%%%%%%%

% Don't change these lines
\bsp	% typesetting comment
\label{lastpage}
\end{document}